# Near-Field Enhancement via Plasmon–Phonon Polariton Coupling with CdO Stripes

P. Ibañez-Romero[1], Maria Villanueva-Blanco[1], Javier Yeste[2], F. Gonzalez-Posada[3], T. Taliercio[3], V. Muñoz-Sanjosé[2], M. Montes Bajo[1*], and A. Hierro[1*]

[1]ISOM, Universidad Politécnica de Madrid, Madrid, Spain

[2]Departament de Física Aplicada i Electromagnetisme, Universitat de València, Burjassot, Spain

[3]Univ. Montpellier, IES. UMR 5214, F-34000, Montpellier, France

Contact: miguel.montes@upm.es, adrian.hierro@upm.es

ABSTRACT

The mid-infrared spectral region presents significant potential for sensing and spectroscopic applications. However, traditional plasmonic materials exhibit substantial optical losses within this range, thereby constraining their effectiveness. Emerging materials such as cadmium oxide (CdO) have demonstrated promise in overcoming these limitations. In this work, we introduce a novel approach to engineer large coupling between localized surface plasmons (LSPs) in CdO and localized surface phonon polaritons (LSPhP) in sapphire. By developing a successful dry etching protocol for CdO, we fabricate stripe arrays with tunable sizes, allowing the spectral alignment between the LSP and LSPhP modes. We demonstrate both experimentally and numerically that when these polaritons become resonant, hybrid modes emerge, resulting in coupling. Finite element simulations reveal near-field enhancements exceeding a factor of 1000, spatially extended hundreds of nanometers around the etched structures. Our approach bridges the plasmonic and phononic responses of two mid-IR active materials, paving the way for scalable, high-performance infrared sensing platforms.

INTRODUCTION

The mid-infrared (mid-IR) region of the electromagnetic spectrum has consistently drawn interest due to its potential in a wide range of applications, including medical diagnostics, homeland security, and chemical and biological sensing [1], [2], [3], [4], [5]. Of particular relevance is its role in molecular spectroscopy, as it encompasses the so-called *molecular fingerprint region* [6], where the resonant excitation of specific chemical bonds enables precise molecular identification [7], [8]. To fully exploit this spectral window, the field of plasmonics has expanded into the mid-IR regime [9], [10]. However, conventional plasmonic materials such as noble metals suffer at these frequencies from high losses [11] and a limited electric field confinement, which can only be achieved by using ultra small gaps between the metallic surfaces (< 20 nm) [12], making them difficult to manufacture and limiting their effectiveness. This has prompted the search for alternative materials capable of sustaining low-loss, tunable plasmonic responses in the mid-IR [13].

Among the emerging candidates in the mid-IR, cadmium oxide (CdO) stands out as a leading plasmonic material in the spectral range from 1 to 10 μm. As demonstrated by Caldwell *et al.* [14], CdO exhibits the highest figure of merit for localized surface plasmons (LSPs) in this spectral region. Its large screened plasma frequency, $\omega_p$, tunable up to 4000 cm$^{-1}$ or 2.5 μm, combined with low optical losses, γ typically 550 cm$^{-1}$ [15], enables strong, plasmonic resonances suitable for a variety of mid-IR applications [16], [17], [18].

Traditionally, r-plane sapphire has been the substrate of choice for the growth of high quality CdO, as it provides excellent crystallographic alignment due to their closely matched lattice parameters [19], [20]. In addition, sapphire has long been employed as a substrate in solid-state technologies owing to its wide bandgap, large electrical resistivity, and high thermal conductivity. Due to sapphire's uniaxial optical anisotropy [21], it supports multiple IR-active phonon modes in the mid-IR, four with dipole moments perpendicular (planar) and two parallel (axial) to the c-axis, spanning from 385 cm$^{-1}$ to 907 cm$^{-1}$. These modes enable the generation of surface phonon polaritons with high confinement and low losses, making sapphire an outstanding platform for phononic applications [18], [22], [23]. All in all, sapphire and CdO provide an outstanding platform for proving plasmon-phonon coupling in the mid-IR. However, a critical challenge remains: the plasma frequency of CdO and the phonon-active range of sapphire do not naturally overlap, hindering efficient coupling between their respective resonant modes. Previous attempts to address this mismatch have resorted to doping [24], [25], [26] or alloying CdO [27] in order to tune its plasma frequency. However these approaches achieve a $\omega_p$ spectrally distant from the active phonon range in sapphire, thus the coupling between both oscillators is negligible. Lastly, other strategies have focused on the bottom-up synthesis of nanoscale geometries, aiming to place the LSPs at desired spectral positions [29], [30], [31], [32]. However, there has not been any successful reports yet on the technological process of CdO layers to obtain nanostructures of arbitrary and controlled geometry.

Here, we report for the first time the processing of CdO layers to tune the LSP resonances at will, bridging the spectral gap between CdO plasmons and sapphire phonons. By etching stripes into CdO thin films grown on r-plane sapphire, we induce two key effects: the formation of tunable localized surface plasmons in CdO and the activation of localized surface phonon polaritons (LSPhPs) within the Reststrahlen band of sapphire. As the spacing between stripes is reduced and the width of the stripes is increased, the LSP resonance of CdO shifts toward lower frequencies. When the LSP and LSPhP modes are spectrally aligned, they couple, giving rise to hybrid modes with distinct signatures in both the far and near fields. In this work, we present experimental evidence of this coupling on the far field, alongside computational results that highlight their significance in the near field. Using a finite element method (FEM) model, we demonstrate that this coupling can enhance the local electromagnetic field by more than a factor of 1000, on par with the state of the art [33], [34], [35], [36].

RESULTS

In this work, the mid-IR plasmonic response of CdO is tailored by controlling the surface morphology. Stripes of various sizes were formed on CdO grown on sapphire through the process illustrated in Figure 1a. It begins with the MOCVD deposition of CdO films on top of r-plane sapphire wafers. Then, standard e-beam lithography is used to define stripes on a 700 nm layer of PMMA resist. After development, the pattern is transferred to the CdO film using a novel dry etching protocol developed in house: 30 sccm of $BCl_3$ and 10 sccm of Ar are fed into an inductive coupled plasma reactive ion etching (ICP RIE) chamber and the power is set to 250 W, keeping the pressure between 10 and 20 mTorr. The etching speed is around 1 nm s$^{-1}$ so, depending on the thickness of the CdO film, the etching time is set accordingly. This protocol was also successfully applied on CdZnO samples as discussed in the supporting information.

The resulting surface is shown in Figure 1b, where scanning electron micrographs of three samples are presented. The design of the samples consists of 200 x 200 µm$^2$ square areas where the stripes are defined. The design parameters are the period (d), which is kept constant at 2 µm for all the samples shown in this study, and the gap between stripes (w). Both parameters are shown in figure 1b. To simplify matters, we use the fill factor $FF = \left(1 - \frac{w}{d}\right)$, to describe the stripe geometry. In this work, a series of samples with FF ranging from 0.3 to 0.96 were fabricated to explore the tunability of the system. Figure 1c shows the AFM profile of the sample with FF=0.7. The stripes have a cross-sectional trapezoidal shape and both the top surface of the CdO and the surface of the sapphire between the stripes are flat within 7 RMS nm. Notably the processing protocol allows for a separation between the stripes as small as 80 nm at the sapphire-CdO plane.

The resulting surface morphology together with the plasmonic properties of CdO and the polar character of sapphire give rise to the appearance of characteristic features in the mid-IR. For both types of materials (polar dielectrics and semiconductors) polaritonic excitations occur when the dielectric permittivity of the surrounding medium is of opposite sign to that of their own. When these excitations are localized they are called LSPs in the case of semiconductors, and for polar dielectrics LSPhP. As shown in Figure 1d, there are two simplified scenarios in the set up presented in this work. The first scenario (top) represents the case when the permittivity of sapphire is positive (frequency above its Reststrahlen band), but that of CdO is negative (frequency below its plasma frequency). This allows the appearance of a LSP in the CdO-air interface. The second scenario (bottom) occurs when both the permittivities of the substrate and the stripes are negative. This allows for the appearance of LSPhP in the sapphire-air interfaces as well as LSP in the CdO-air interface. It is in the latter scenario that by modifying the FF during the processing of the samples, both excitonic responses can be made resonant taking advantage of the coupling. However, for the sake of clarity, let's first analyze both features separately.

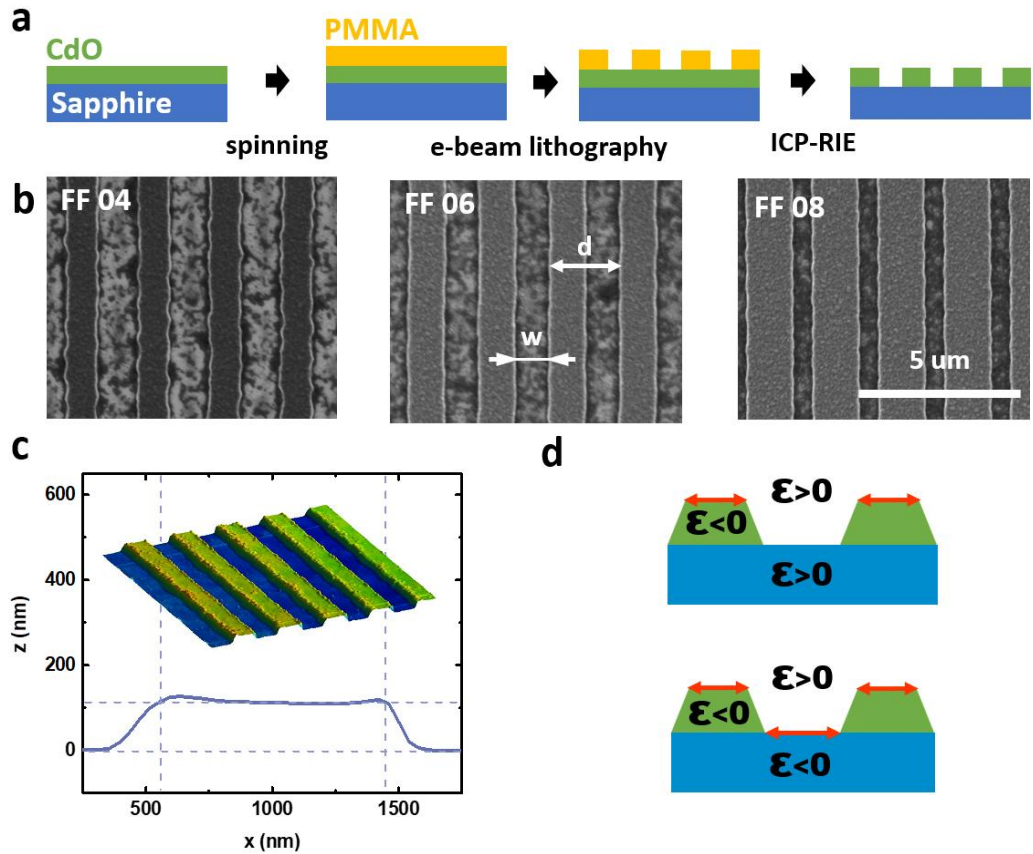

Figure 1. (a) Processing scheme of the samples. (b) SEM micrographs for three different FF, from left to right: 0.4, 0.6 and 0.8. (c) AFM profile of stripes with FF=0.7. (d) Sketch of LSP and LSPhP formation for two different scenarios: top when only the LSP is formed; bottom, when both an LSP and LSPhP are formed.

As an initial step, we examine the substrate's response, as it plays a crucial role in the observed phononic behavior. Sapphire, being anisotropic, supports both planar (E) and axial (A) phonons. The growth of high quality CdO requires the use of r-plane sapphire, with its c-axis tilted approximately 58° from the surface normal, which results in the simultaneous excitation of both phonon types when measured at normal incidence. To better describe the underlying mechanisms while maintaining computational efficiency, a simplified yet accurate enough finite element methods (FEM) model was developed considering only the planar phonons of sapphire. Details and validation of this model are provided in Figure 1S of the Supporting Information.

Polar materials like sapphire can sustain LSPhPs within their Reststrahlen band [37], a spectral region characterized by high reflectivity. These excitations are governed by the shape of the nanostructure rather than its absolute dimensions, provided the system operates below the diffraction limit [14]. As a result, increasing the FF of the design does not shift the LSPhP resonance, which remains spectrally pinned. This is in clear contrast to the behavior of the LSPs supported by the CdO stripes, which are highly sensitive to the geometric parameters of the structure that hosts them. LSPs are induced by the shape of the stripes, giving rise to plasma oscillations in the direction transversal to the longitudinal axis of the stripes when measured at normal incidence with the electric field perpendicular to the stripes. Such behavior allows for fine tuning of the spectral position of the LSP peak resonance by changing the FF. Figure 2a shows the reflectance at 18° incidence (see Methods) for stripes with FF from 0.3 to 0.8, achieving a shift of the peak resonance from 1780 to 1230 $cm^{-1}$. The intensity of the reflectance peak decreases for smaller FF because the area covered by CdO decreases, which ultimately means less electrons can contribute to the collective oscillations. Figure 2a also shows the fine agreement between the measured spectra and the calculated curves for all the different FFs.

Upon close inspection of the spectra, the LSPs present an asymmetric shape with a slight shoulder at higher frequencies. The origin of such feature is linked to higher order excitations that occur in the stripes as shown in Figure 2b, where the vertical component of the electric field is plotted at its highest average value. The field distribution of the secondary mode is different from that of the primary mode at the lower corner of the slanted wall. The overall intensity of the field is also lower as the secondary mode is less intense than the main one.

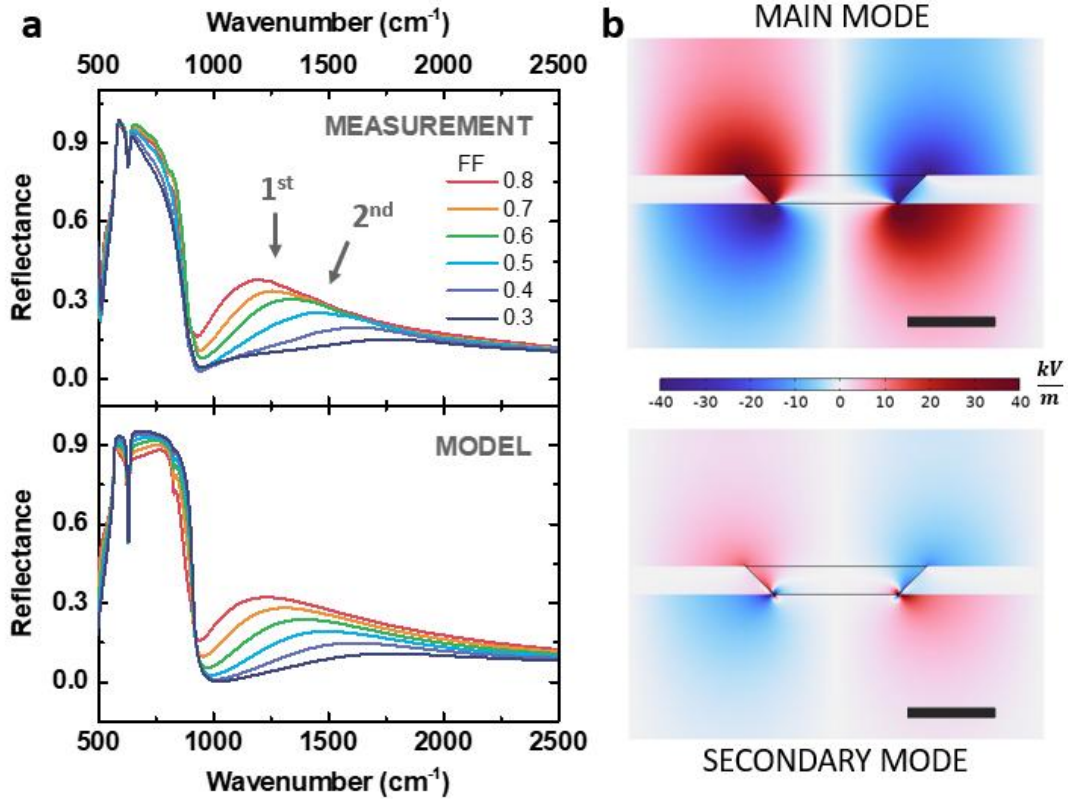

Figure 2. (a) Experimental and calculated reflectance spectra of stripes with different FF measured through a microscope lens. (b) Distribution of the vertical component of the electric field at the main mode and the secondary mode peak frequency. The black bar represents a distance of 500 nm.

However, as the FF increases beyond 0.8 a new scenario arises. In a hypothetical scenario where the substrate does not support phonons, the LSP resonance could continuously be tuned to lower frequencies by adjusting the stripe geometry, eventually bypassing the Reststrahlen band associated with sapphire's fourth planar phonon, as illustrated in Figure S2. In the current system, however, the Reststrahlen band is present and supports a LSPhP, which can observed in Figure 3a and 3b as a subtle dip in the reflectance spectrum near 830 cm$^{-1}$ for a FF of 0.78. As previously noted, the LSPhP remains spectrally fixed across all geometries. In contrast, the LSP resonance is geometry-dependent and shifts with increasing FF. This tunability enables progressive spectral matching between the LSP and LSPhP, leading to their coupling. This coupling becomes evident in the reflectance data presented in Figure 3a, which is reproduced with excellent agreement by the numerical results obtained with the model and shown in Figure 3b. The reflectance at the LSPhP frequency drops from approximately 0.8 at FF = 0.78

(uncoupled regime) to around 0.4 at FF = 0.96, where the LSP and LSPhP are spectrally matched. The evolution of this coupling is more clearly illustrated in the absorptance spectra derived from the FEM model (Figure 3d). As the LSP approaches the phonon resonance, it initially enhances the absorption associated with the $E_4$ longitudinal phonon. Once the LSP fully overlaps with the LSPhP, the absorptance increases significantly inside of the Reststrahlen band, rising from 0.2 in the uncoupled case to 0.8 in the coupled regime. Figure 3c compiles the absorptance spectra providing a clear visualization of the spectral evolution. As the LSP enters the Reststrahlen band, a pronounced absorption peak emerges at the LSPhP frequency, confirming the onset of coupling. Additionally, a secondary absorption feature appears between the second and third Reststrahlen bands. This is attributed to the low-energy tail of the LSP, visible as a shoulder in the reflectance spectra (Figures 4a and 4b) and as a weak absorption peak that shifts to lower frequencies with increasing FF in Figure 3d. To support this interpretation, Figure 3c also includes the calculated spectral positions of the LSP in the absence of substrate phonons, extracted from Figure S2 of the supporting information. These calculations confirm that the LSP can indeed be tuned across the LSPhP frequency purely through geometric modification, validating the central premise that coupling between the two resonances is achieved by tuning the LSP via the FF.

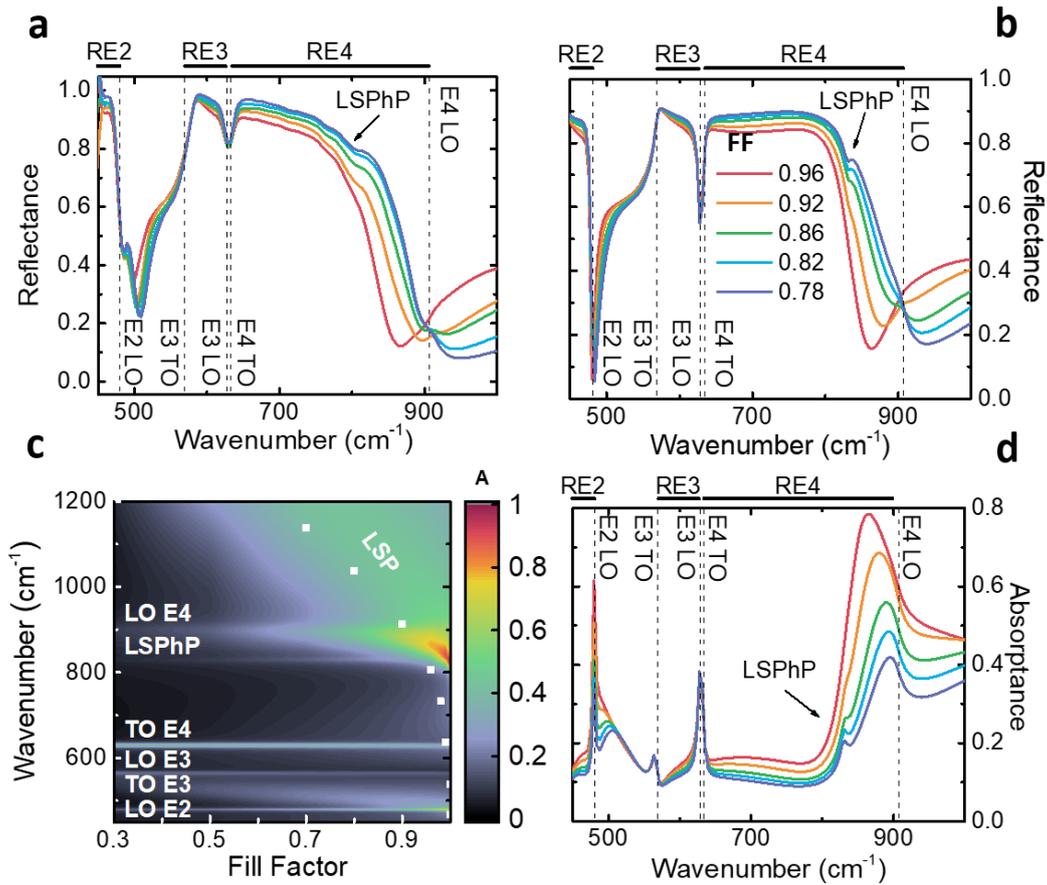

Figure 3. Calculated (a) and experimental (b) reflectance spectra for stripes with FF between 0.78 and 0.96. Calculated absorptance spectra (c) and absorptance dispersion with FF (d). The white dots represent the calculated wavenumber of the LSP in the absence of coupling (details in the Supplemental Information).

In terms of the near field, the coupling between the LSP and the LSPhP has outstanding effects. Figure 4a shows the field enhancement in the surroundings of the stripes for a FF of 0.96 (fully coupled) at the resonance frequency (830 cm$^{-1}$). The light intensity enhancement maps are calculated as follows: the values for $|E_{st}|^2$ are obtained from the model that includes the stripes, while the values for $|E_0|^2$ come from a simulation without them. By comparing both field distributions, the enhancement ($|E_{st}|^2/|E_0|^2$) is determined. When the LSP and LSPhP are fully coupled, the intensity of the field at the lower corner of the slanted walls is greatly enhanced, surpassing values of 1000 at some points inside the area defined by the walls of the stripes (which will be addressed as the hotspot area for the rest of the manuscript). This value is ahead from previous studies with CdO nanostructures [29], and compares well with the state of the art [33], [34], [35], [36]. Furthermore, the area at which the electric field is magnified is high and wide, reaching a tenfold enhancement at 150 nm above the surface of the nanostripes, 200 nm into the substrate and a 100 nm to the sides of the upper corner of the walls. Moreover, the field is amplified in air up to 700 nm on top of the nanostripe, and 500 nm laterally.

Figure 4b shows a simulation carried out taking away the phonon resonances from the permittivity function of the substrate and only keeping its high frequency dielectric constant, thus removing the LSPhP. As can be seen in Figure 4b, the overall enhancement in this scenario is much lower, reaching a maximum value of 40 at the lower corner of the slanted walls, only a 4% of the maximum enhancement achieved when both phonons and plasmons are present and coupled. The spread of the enhancement is also greatly reduced, falling quickly below the tenfold values within the hotspot area. The enhancement has also been compared spectrally between the polar and non-polar cases in Figure 4c. To obtain both curves, the enhancement was averaged inside the hotspot area as a function of frequency around the Reststrahlen band [38]. In the polar substrate case, the enhancement is maximum —over 200— at the coupling frequency decreasing rapidly for other frequencies. In the non-polar substrate case the averaged enhancement is generally lower —staying always below 40— and it monotonically increases towards lower energies. These findings unequivocally support the idea of coupling between the polar response of sapphire and the plasmonic response of CdO.

Further, the same approach was used to visualize the progressive coupling of the LSP and the LSPhP in the near field for growing FF in Figure 4d. When the FF is low (0.3) the LSP occurs at much higher energies than the LSPhP, which is visible at 830 cm$^{-1}$, effectively preventing their coupling (see Figure S4). At this stage the surface-averaged enhancement in the hotspot area of the LSP is approximately 2.6, while that of the LSPhP is around 3.5. As the stripes are brought closer together, the near-field enhancement intensifies around the LSPhP frequency. This enhancement reaches a peak value of 236 when both excitations are fully coupled. In other words, coupling between the LSP and LSPhP leads to a field enhancement in the hotspot region that is around 67 times greater than when the excitations remain uncoupled. Additionally, the geometry of the system may offer practical advantages for specific applications. For example, the tilted sidewalls of the structure improve analyte access to the hotspot regions, which can significantly enhance the performance of surface-enhanced infrared absorption (SEIRA) spectroscopy.

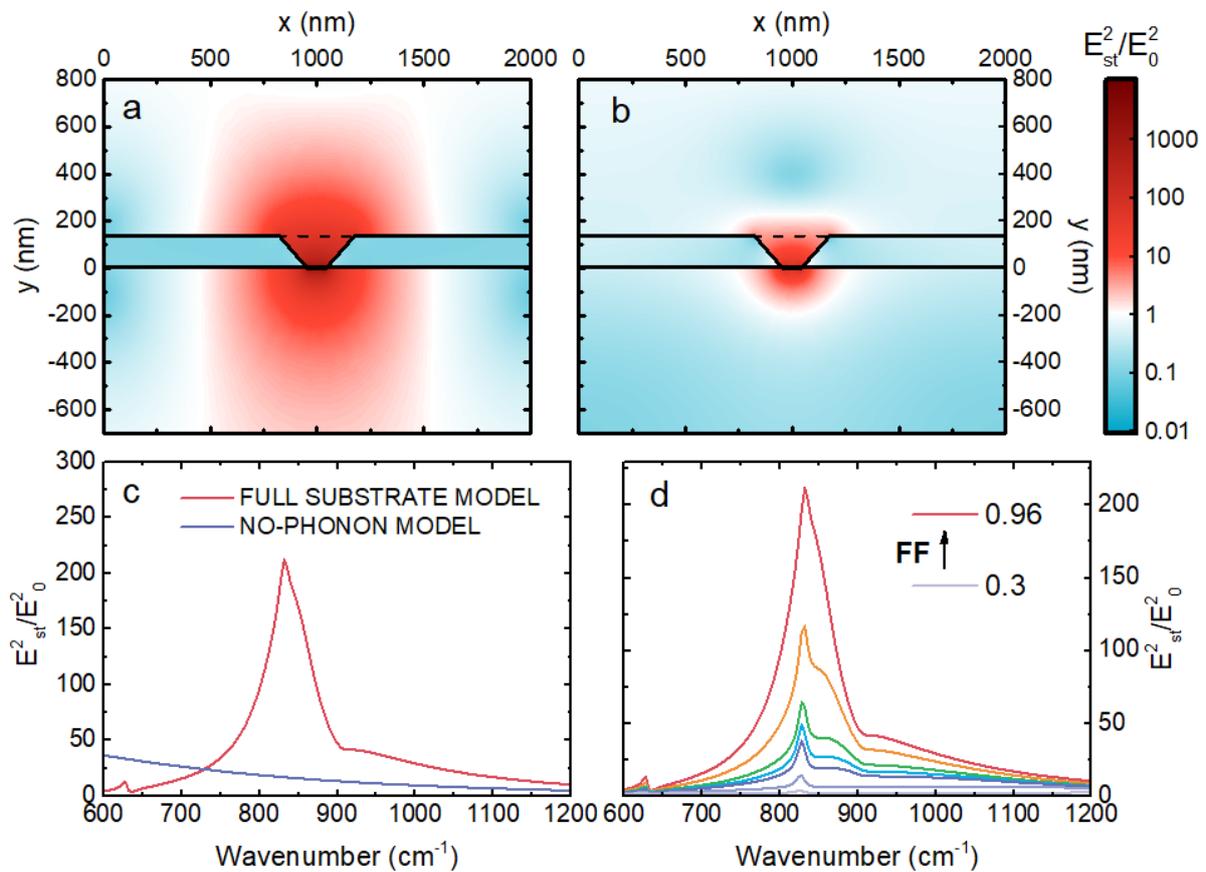

Figure 4. Intensity enhancement maps at 830 cm$^{-1}$ for stripes with full substrate model (a), and no phonon substrate model (b). The black lines define the geometry of the stripes and the dotted lines are shown to mark the space deemed as 'hotspot' area in the text. The surface-averaged field enhancement for the full model of the sapphire substrate and the no-phonon sapphire model is shown in (c), and for different stripe FF with the full model in (d).

CONCLUSIONS

In this work, we have demonstrated a novel approach for engineering the coupling between LSPs in CdO and LSPhPs in r-plane sapphire within the mid-infrared spectral region. By developing and implementing the first successful dry etching protocol for CdO, we have enabled the fabrication of stripe arrays with precisely controlled geometries. This processing technique allows to tune the LSP resonances on CdO simply by varying the FF of the patterned stripes, ultimately enabling spectral alignment with the LSPhP supported by the sapphire substrate. Such alignment results in the emergence of hybrid modes, as evidenced by both experimental reflectance measurements and detailed finite element simulations. The observed coupling is manifested by pronounced changes in both far-field reflectance and absorptance, with the absorption at the LSPhP frequency highly increasing as the system transitions into the coupled regime. Most notably, our computational studies reveal a significant enhancement of the local electromagnetic field in the vicinity of the nanostructures. When the LSP and LSPhP are fully coupled, the near-field intensity enhancement exceeds a factor of 1000, on par with the state of the art. This enhancement is not only exceptionally high but also spatially broad, extending hundreds of nanometers from the stripe surfaces.

## METHODS

### Mid-IR spectroscopy

Infrared spectra were taken using a Fourier Transform Infrared (FTIR) spectrometer equipped with a nitrogen cooled mercury cadmium telluride (MCT) detector, a KBr beamsplitter, a SiC glowbar and a IR Schwartzschild microscope fitted with a x 36 objective. The samples were measured in a reflectance configuration with an incident cone of angles between 12 a 24º, yielding an average of 18º, and light polarized perpendicular to the stripes' longitudinal axis. A gold mirror served as reference for normalization of the spectrum acquired on the sample and all spectra were recorded with a resolution of 4 cm$^{-1}$ and 1,024 iterations.

### Thin film growth

The CdO thin films featured in this work were grown by MOCVD on single-side polished r-plane sapphire, using tertiary butanol and dimethylcadmium, with $N_2$ as the carrier gas, with the temperature kept at 276 ºC during 7 minutes. The choice of substrate plane was done because r-plane sapphire yields the highest quality films due to the reduced mismatch between the CdO layer and the substrate [39], [40]. Moreover, sapphire has a wide band gap, which ensures high transparency across a broad frequency range. Its large thermal conductivity enhances sample homogeneity by enabling uniform temperature distribution throughout the sample. All in all, CdO films grown on r-sapphire exhibit excellent crystallinity and flat surfaces, which facilitates the dry etch during processing. The final film thickness was 137 nm, which was confirmed with AFM after etching.

Finite elements method model

A 2D finite element model was created in COMSOL Multiphysics to analyze the electric field distribution around the stripes. The model features a sapphire substrate on top of which the CdO stripes are defined and the resulting structures are covered by air. The chosen geometrical parameters were adapted from the AFM and SEM results obtained on the stripes samples. With respect to the materials, the plasmonic response of CdO in the mid-IR was modeled through Drude's equation:

$$\varepsilon(\omega) = \varepsilon_\infty^{CdO} - \frac{\omega_p^2}{\omega^2 + i\omega\gamma_p}$$

featuring a plasma frequency ($\omega_p$), its associated damping ($\gamma_p$) and the high-frequency dielectric constant ($\varepsilon_\infty$). The values for these parameters were obtained though fitting the reflectance spectra of the unprocessed CdO films. Sapphire's phononic response was defined through the Gervais model:

$$\varepsilon(\omega) = \varepsilon_\infty^{Sapp} \prod_{i=1} \frac{\omega_{LO,i}^2 - \omega^2 - i\omega\gamma_{LO,i}}{\omega_{TO,i}^2 - \omega^2 - i\omega\gamma_{TO,i}}$$

as a product of harmonic oscillators for each phonon pair (LO/TO) with its associated frequency ($\omega_{TO,i}$, $\omega_{LO,i}$) and damping ($\gamma_{TO,i}$, $\gamma_{LO,i}$). Such values were extracted from Schubert et al [21]. The parameter values are shown in Table 1.

The model uses linearly polarized light to emulate the incident beam of the FTIR spectrometer. The electric field is polarized in TM configuration and has an incidence angle of 18º (averaged from the angle range of the Scharzschild objective).

Table 1. Modelling parameters of the materials' dielectric response

| Material | CdO | Sapphire |
|---|---|---|
| $\varepsilon_\infty$ | 5.3 | 3.1 |
| $\omega_p, \gamma_p$ (cm$^{-1}$) | 6428, 523 | |
| $\omega_{TO1}, \gamma_{TO1}$ (cm$^{-1}$) | - | 385, 3.1 |
| $\omega_{LO1}, \gamma_{LO1}$ (cm$^{-1}$) | - | 388, 3.1 |
| $\omega_{TO2}, \gamma_{TO2}$ (cm$^{-1}$) | - | 439, 3.1 |
| $\omega_{LO2}, \gamma_{LO2}$ (cm$^{-1}$) | - | 481, 1.9 |
| $\omega_{TO3}, \gamma_{TO3}$ (cm$^{-1}$) | - | 569, 4.7 |
| $\omega_{LO3}, \gamma_{LO3}$ (cm$^{-1}$) | - | 629, 5.9 |
| $\omega_{TO4}, \gamma_{TO4}$ (cm$^{-1}$) | - | 634, 5.0 |
| $\omega_{LO4}, \gamma_{LO4}$ (cm$^{-1}$) | - | 907, 14.7 |

ACKNOWLEDGEMENTS

This work was partly supported by Project PID2024-156706OB-C2 funded by MICIU/AEI/10.13039/501100011033/; PID2020-114796RB-C2 funded by MCIN/ AEI /10.13039/501100011033/; Project PDC2023-145827-C2 funded by MICIU/AEI /10.13039/501100011033/ and by European Union Next Generation EU/ PRTR; Generalitat Valenciana under Project No. PROMETEU/2021/066; and by the projects SEA (Occitanie French Region – ESR-PREMAT-238), EXTRA (ANR 11-EQPX-0016), NanoElastir (ASTRID 2020–2023), and ENVIRODISORDERS (MUSE UM 2021–2023). PI and JY acknowledge the FPU (Formación de Profesorado Universitario) predoctoral contract from the Spanish Ministry of Science, Innovation, and Universities (MICIU). The authors also acknowledge the support provided by the MICRONANOFABS ICTS network from the MICIU.

STATEMENT OF DATA SHARING

The data that support the findings of this study are available from the corresponding author upon reasonable request.

STATEMENT OF CONFLICT OF INTEREST

No potential competing interest was reported by the author(s).

REFERENCES

[1] F. Neubrech, C. Huck, K. Weber, A. Pucci, and H. Giessen, "Surface-Enhanced Infrared Spectroscopy Using Resonant Nanoantennas," *Chem. Rev.*, vol. 117, no. 7, pp. 5110–5145, Apr. 2017, doi: 10.1021/acs.chemrev.6b00743.

[2] R. Soref, "Mid-infrared photonics in silicon and germanium," *Nature Photon*, vol. 4, no. 8, pp. 495–497, Aug. 2010, doi: 10.1038/nphoton.2010.171.

[3] K. Ataka, T. Kottke, and J. Heberle, "Thinner, Smaller, Faster: IR Techniques To Probe the Functionality of Biological and Biomimetic Systems," *Angew Chem Int Ed*, vol. 49, no. 32, pp. 5416–5424, Jul. 2010, doi: 10.1002/anie.200907114.

[4] T. Taliercio et al., "Fano-like resonances sustained by Si doped InAsSb plasmonic resonators integrated in GaSb matrix," *Opt. Express*, vol. 23, no. 23, p. 29423, Nov. 2015, doi: 10.1364/OE.23.029423.

[5] H.-L. Wang, E.-M. You, R. Panneerselvam, S.-Y. Ding, and Z.-Q. Tian, "Advances of surface-enhanced Raman and IR spectroscopies: from nano/microstructures to macro-optical design," *Light Sci Appl*, vol. 10, no. 1, p. 161, Aug. 2021, doi: 10.1038/s41377-021-00599-2.

[6] L. Maidment, P. G. Schunemann, and D. T. Reid, "Molecular fingerprint-region spectroscopy from 5 to 12 μm using an orientation-patterned gallium phosphide optical parametric oscillator," *Opt. Lett., OL*, vol. 41, no. 18, pp. 4261–4264, Sep. 2016, doi: 10.1364/OL.41.004261.

[7] S. Prucnal et al., "Ultra-doped n-type germanium thin films for sensing in the mid-infrared," *Sci Rep*, vol. 6, no. 1, p. 27643, Jun. 2016, doi: 10.1038/srep27643.

[8] A. Hartstein, J. R. Kirtley, and J. C. Tsang, "Enhancement of the Infrared Absorption from Molecular Monolayers with Thin Metal Overlayers," *Phys. Rev. Lett.*, vol. 45, no. 3, pp. 201–204, Jul. 1980, doi: 10.1103/PhysRevLett.45.201.

[9] A. Boltasseva and H. A. Atwater, "Low-Loss Plasmonic Metamaterials," *Science*, vol. 331, no. 6015, pp. 290–291, Jan. 2011.

[10] T. Taliercio and P. Biagioni, "Semiconductor infrared plasmonics," *Nanophotonics*, vol. 8, no. 6, pp. 949–990, Jun. 2019, doi: 10.1515/nanoph-2019-0077.

[11] B. Cerjan, X. Yang, P. Nordlander, and N. J. Halas, "Asymmetric Aluminum Antennas for Self-Calibrating Surface-Enhanced Infrared Absorption Spectroscopy," *ACS Photonics*, vol. 3, no. 3, pp. 354–360, Mar. 2016, doi: 10.1021/acsphotonics.6b00024.

[12] M. Najem, F. Carcenac, T. Taliercio, and F. Gonzalez-Posada, "Aluminum Bowties for Plasmonic-Enhanced Infrared Sensing," *Advanced Optical Materials*, vol. 10, no. 20, p. 2201025, Oct. 2022, doi: 10.1002/adom.202201025.

[13] P. R. West, S. Ishii, G. V. Naik, N. K. Emani, V. M. Shalaev, and A. Boltasseva, "Searching for better plasmonic materials," *Laser & Photon. Rev.*, vol. 4, no. 6, pp. 795–808, Nov. 2010, doi: 10.1002/lpor.200900055.

[14] J. D. Caldwell et al., "Low-loss, infrared and terahertz nanophotonics using surface phonon polaritons," *Nanophotonics*, vol. 4, no. 1, pp. 44–68, Apr. 2015.

[15] E. Martínez Castellano et al., "Mid-IR Surface Plasmon Polaritons in CdZnO thin films on GaAs," *Applied Surface Science*, vol. 608, p. 155060, Jan. 2023, doi: 10.1016/j.apsusc.2022.155060.

[16] J. D. Caldwell, I. Vurgaftman, J. G. Tischler, O. J. Glembocki, J. C. Owrutsky, and T. L. Reinecke, "Atomic-scale photonic hybrids for mid-infrared and terahertz nanophotonics," *Nature Nanotech*, vol. 11, no. 1, pp. 9–15, Jan. 2016, doi: 10.1038/nnano.2015.305.

[17] J. Tamayo-Arriola, A. Huerta-Barberà, M. Montes Bajo, E. Muñoz, V. Muñoz-Sanjosé, and A. Hierro, "Rock-salt CdZnO as a transparent conductive oxide," *Applied Physics Letters*, vol. 113, no. 22, p. 222101, Nov. 2018, doi: 10.1063/1.5048771.

[18] J. D. Caldwell, E. Sachet, C. Shelton, and T. G. Folland, "Filterless non-dispersive infrared sensing devices and methods," US20230221242A1, Jul. 13, 2023 Accessed: Jan. 10, 2024. [Online]. Available: https://patents.google.com/patent/US20230221242A1/en?oq=20230221242

[19] J. Yeste Torregrosa, "MOCVD Growth of Rocksalt Cd(Zn)O: Reactor Characterization and Integration on Functional Substrates for Applications in Mid-IR Plasmonics," 2024, Accessed: Mar. 19, 2025. [Online]. Available: https://hdl.handle.net/10550/104751

[20] J. Zuñiga-Pérez, C. Munuera, C. Ocal, and V. Muñoz-Sanjosé, "Structural analysis of CdO layers grown on r-plane sapphire ( 0 1 1 ¯ 2 ) by metalorganic vapor-phase epitaxy," *Journal of Crystal Growth*, vol. 271, no. 1–2, pp. 223–228, Oct. 2004, doi: 10.1016/j.jcrysgro.2004.07.069.

[21] M. Schubert, *Infrared Ellipsometry on Semiconductor Layer Structures: Phonons, Plasmons, and Polaritons*. in Springer Tracts in Modern Physics, no. 209. Berlin, Heidelberg: Springer Berlin Heidelberg, 2005. doi: 10.1007/b11964.

[22] S. Xie *et al.*, "LSP-SPP Coupling Structure Based on Three-Dimensional Patterned Sapphire Substrate for Surface Enhanced Raman Scattering Sensing," *Nanomaterials*, vol. 13, no. 9, p. 1518, Apr. 2023, doi: 10.3390/nano13091518.

[23] A. Bile *et al.*, "Tuning of the Berreman mode of GaN/AlxGa1-xN heterostructures on sapphire: The role of the 2D-electron gas in the mid-infrared," *Optical Materials*, vol. 147, p. 114708, Jan. 2024, doi: 10.1016/j.optmat.2023.114708.

[24] T. R. Gordon *et al.*, "Shape-Dependent Plasmonic Response and Directed Self-Assembly in a New Semiconductor Building Block, Indium-Doped Cadmium Oxide (ICO)," *Nano Lett.*, vol. 13, no. 6, pp. 2857–2863, Jun. 2013, doi: 10.1021/nl4012003.

[25] E. Przezdziecka *et al.*, "Short-Period CdO/MgO Superlattices as Cubic CdMgO Quasi-Alloys," *Crystal Growth & Design*, vol. 20, no. 8, pp. 5466–5472, Aug. 2020, doi: 10.1021/acs.cgd.0c00678.

[26] A. J. Cleri *et al.*, "Tunable, Homoepitaxial Hyperbolic Metamaterials Enabled by High Mobility CdO," *Advanced Optical Materials*, vol. 11, no. 1, p. 2202137, 2023, doi: 10.1002/adom.202202137.

[27] J. Tamayo Arriola, "Infrared intersubband detection and plasmonics with (Zn,Mg)O and (Cd,Zn)O compounds," PhD Thesis, Universidad Politécnica de Madrid, 2019. doi: 10.20868/UPM.thesis.57422.

[28] J. Tamayo-Arriola *et al.*, "Controllable and Highly Propagative Hybrid Surface Plasmon–Phonon Polariton in a CdZnO-Based Two-Interface System," *ACS Photonics*, vol. 6, no. 11, pp. 2816–2822, Nov. 2019, doi: 10.1021/acsphotonics.9b00912.

[29] P. Ibañez-Romero *et al.*, "Large-area sensors using Cd(Zn)O plasmonic nanoparticles for surface-enhanced infrared absorption," *Nanophotonics*, May 2025, doi: 10.1515/nanoph-2025-0020.

[30] A. Tadjarodi and M. Imani, "A novel nanostructure of cadmium oxide synthesized by mechanochemical method," *Materials Research Bulletin*, vol. 46, no. 11, pp. 1949–1954, Nov. 2011, doi: 10.1016/j.materresbull.2011.07.016.

[31] T. Ghoshal, S. Biswas, P. M. G. Nambissan, G. Majumdar, and S. K. De, "Cadmium Oxide Octahedrons and Nanowires on the Micro-Octahedrons: A Simple Solvothermal Synthesis," *Crystal Growth & Design*, vol. 9, no. 3, pp. 1287–1292, Mar. 2009, doi: 10.1021/cg800203y.

[32] D. Z. Husein, R. Hassanien, and M. Khamis, "Cadmium oxide nanoparticles/graphene composite: synthesis, theoretical insights into reactivity and adsorption study," *RSC Adv.*, vol. 11, no. 43, pp. 27027–27041, 2021, doi: 10.1039/D1RA04754J.

[33] F. Neubrech, A. Pucci, T. W. Cornelius, S. Karim, A. García-Etxarri, and J. Aizpurua, "Resonant Plasmonic and Vibrational Coupling in a Tailored Nanoantenna for Infrared Detection," *Phys. Rev. Lett.*, vol. 101, no. 15, p. 157403, Oct. 2008, doi: 10.1103/PhysRevLett.101.157403.

[34] H. Aouani *et al.*, "Ultrasensitive Broadband Probing of Molecular Vibrational Modes with Multifrequency Optical Antennas," *ACS Nano*, vol. 7, no. 1, pp. 669–675, Jan. 2013, doi: 10.1021/nn304860t.

[35] L. V. Brown, K. Zhao, N. King, H. Sobhani, P. Nordlander, and N. J. Halas, "Surface-Enhanced Infrared Absorption Using Individual Cross Antennas Tailored to Chemical Moieties," *J. Am. Chem. Soc.*, vol. 135, no. 9, pp. 3688–3695, Mar. 2013, doi: 10.1021/ja312694g.

[36] C. Huck *et al.*, "Surface-Enhanced Infrared Spectroscopy Using Nanometer-Sized Gaps," *ACS Nano*, vol. 8, no. 5, pp. 4908–4914, May 2014, doi: 10.1021/nn500903v.

[37] J. D. Caldwell *et al.*, "Low-Loss, Extreme Subdiffraction Photon Confinement via Silicon Carbide Localized Surface Phonon Polariton Resonators," *Nano Lett.*, vol. 13, no. 8, pp. 3690–3697, Aug. 2013.

[38] K. L. Kelly, E. Coronado, L. L. Zhao, and G. C. Schatz, "The Optical Properties of Metal Nanoparticles: The Influence of Size, Shape, and Dielectric Environment," *J. Phys. Chem. B*, vol. 107, no. 3, pp. 668–677, Jan. 2003, doi: 10.1021/jp026731y.

[39] J. Zúñiga-Pérez, C. Martínez-Tomás, and V. Muñoz-Sanjosé, "X-ray characterization of CdO thin films grown on a-, c-, r- and m-plane sapphire by metalorganic vapour phase-epitaxy," *physica status solidi (c)*, vol. 2, no. 3, pp. 1233–1238, 2005, doi: 10.1002/pssc.200460672.

[40] A. Huerta-Barberà *et al.*, "MOCVD growth of CdO very thin films: Problems and ways of solution," *Applied Surface Science*, vol. 385, pp. 209–215, Nov. 2016, doi: 10.1016/j.apsusc.2016.05.113.